# Three-dimensional particle tracking velocimetry using shallow neural network for real-time analysis

Yeonghyeon Gim,[1] Dong Kyu Jang,[1] Dong Kee Sohn,[1] Hyoungsoo Kim,[2,*] Han Seo Ko[1,*]

[1]School of Mechanical Engineering, Sungkyunkwan University, 2066 Seobu-ro, Jangan-gu, Suwon, Gyeonggi-do 16419, South Korea
[2]School of Mechanical Engineering, KAIST, Daehak-ro 291, Yuseong-gu, Daejeon 34141, South Korea
*Corresponding author: hshk@kaist.ac.kr, hanseoko@skku.edu

**Three-dimensional particle tracking velocimetry (3D-PTV) technique is widely used to acquire the complicated trajectories of particles and flow fields. It is known that the accuracy of 3D-PTV depends on the mapping function to reconstruct three-dimensional particles locations. The mapping function becomes more complicated if the number of cameras is increased and there is a liquid-vapor interface, which crucially affect the total computation time. In this paper, using a shallow neural network model (SNN), we dramatically decrease the computation time with a high accuracy to successfully reconstruct the three-dimensional particle positions, which can be used for real-time particle detection for 3D-PTV. The developed technique is verified by numerical simulations and applied to measure a complex solutal Marangoni flow patterns inside a binary mixture droplet.**

## 1. Introduction

Three-dimensional flow field measurement techniques have been developed for obtaining massive volumetric data sets for high temporal and spatial resolution from high-speed and multiple cameras while simultaneously reducing computation time and increasing reliability **[1-3]**. Among various three-dimensional flow field measurement techniques, it is known that particle tracking velocimetry (PTV) is able to obtain more complicated trajectories of particles and velocity vectors, for instance three-dimensional Reyleigh-Benard convection flow of the laminar-turbulent transition regime **[4]**, three-dimensional vertical flow structures in a submerged hydraulic jump **[5]**, and free-stream flow of the trisonic wind tunnel **[6]**.

A recent review paper for three-dimensional flow field measurement techniques noted that typically for a successful three-dimensional flow field measurement more than three cameras are required **[7]**. Undoubtedly, the computation load is increased with the amount of data set recorded by multiple cameras. On the other hand, although one- or two-camera system was developed for tomographic particle image velocimetry (PIV), to provide a high accuracy, one of the practical issues is to optimize computation time **[8]**. If a conventional mapping function shape is used, it is still limited to reduce actual computation time.

To enhance the speed of the computation time with a high reliability of particle reconstruction, several algorithms have been suggested, such as neural-network PIV using convolutional neural networks (CNNs) and fully connected neural networks (FCNNs). They applied the neural network algorithm to image analysis, improving the computation time and efficiency **[9, 10, 11]**. Genetic algorithm PTV has been widely used to investigate the flow pattern and measure the velocity. This method based on stereoscopic particle pairing is applied to the epipolar line proximity analysis. This method enables to improve matching accuracy of particle pairs between two stereoscopic particle images compared to the conventional particle pairing method **[12]**. Ohmi et al. studied optimization of the epipolar proximity analysis using self-organizing map algorithm, which is based on Kohonen learning known as one of the machine learning algorithms to evaluate the exact three-dimensional coordinates of particles, increasing the recovery rate of particle pairs **[13]**. Recently, another approach to PTV has been introduced by Zhang et al. They used Voroni diagram that is a method for dividing an area into a number of regions distinguishing particle locations. They showed that this method is useful to avoid such spurious particle matching from PTV algorithm. In their study, a single parameter, the radius of each Voronoi cell, is used to measure the unsteady flow such as the pulsating flow that circulates through a mammal heart **[14]**. Schroder et al. used another iterative method using shake the box algorithm to reconstruct the flow field with 4D-PTV. The STB (Shake The Box) algorithm is highly iterative method which couples the IPR (Iterative Particle Reconstruction) method and the OTF (Optical Transfer Function).

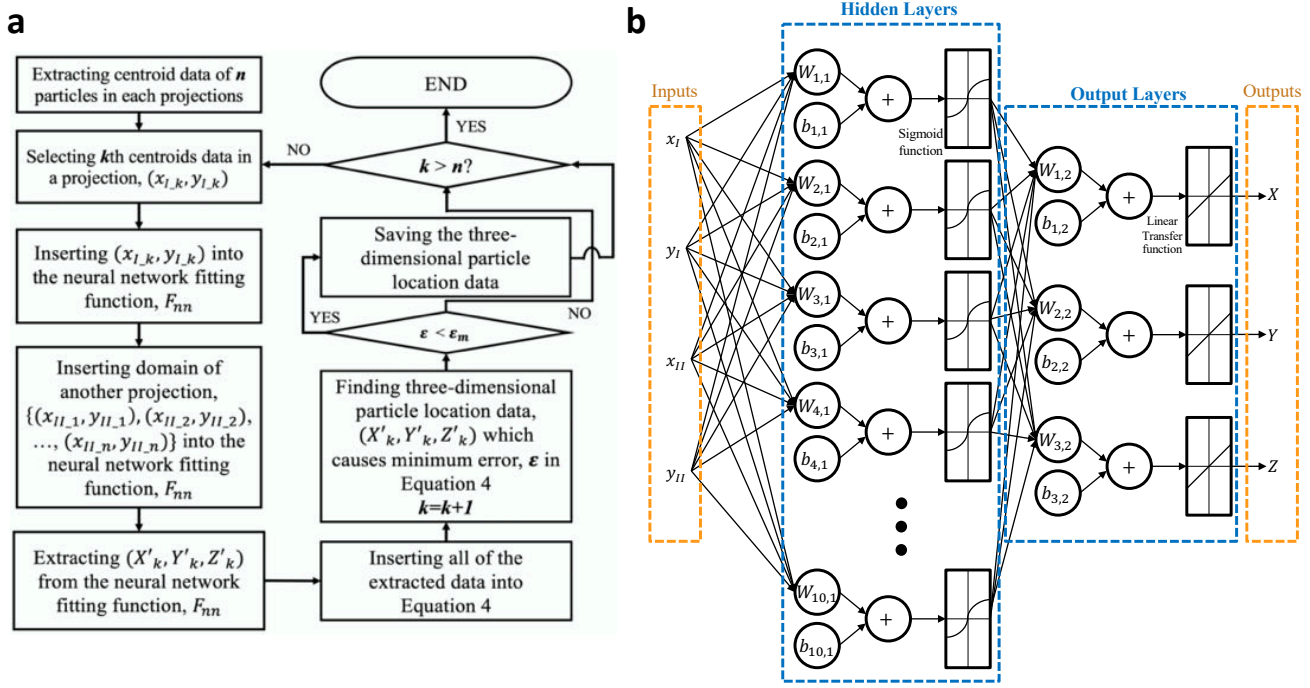

Figure 1. (a) Suggested particle detection algorithm based on shallow neural network (SNN) model where $n$ is the total number of particles and $k$ is an arbitrary particle, $1 \leq k \leq n$. (b) Structure of the neural network system where 10 hidden layers are used.

They reconstructed the highly turbulent region with 6 cameras and investigated spatial and temporal dynamics of structures and particle tracks [15].

It is true that the time for data processing will increase with the amount of data. Nevertheless, among previous studies, few methods considered reducing computing power [9, 14] but these methods do not dramatically decrease the computation time because those still used a conventional mapping function to reconstruct the three-dimensional particle location for PTV. In particular, the mapping function becomes more complicated if there is a refractive index jump due to the liquid-vapor interface, e.g. measuring micro flow inside a liquid drop. In this situation, optical correction is essential, such as a ray tracing technique [16-18] or a shape function [19]. Typically, it is difficult to know the exact shape function of liquid-vapor interfaces, particularly if the shape changes in time. Thus, a complex mapping function approach is widely used and applied to measure three-dimensional flow without the shape function [18]. For the complex mapping function, an indirect iterative method is used, such as gradient descent method [20] or tomography techniques [17], which is conventional but consumes a relatively large amount of the computation time.

In this paper, to minimize the computation time for the reconstruction of three-dimensional particle position, we introduce SNN (shallow neural network)-PTV algorithm, which enables to successfully reconstruct 1000 particles positions in less than 1 second with 99.9 % reconstruction accuracy. We expect that the suggested reconstruction method can be utilized for a real-time flow field measurement. To test our algorithm, we performed numerical simulations and measured the solutal Marangoni flows in the evaporating droplet, which is totally three-dimensional flow pattern.

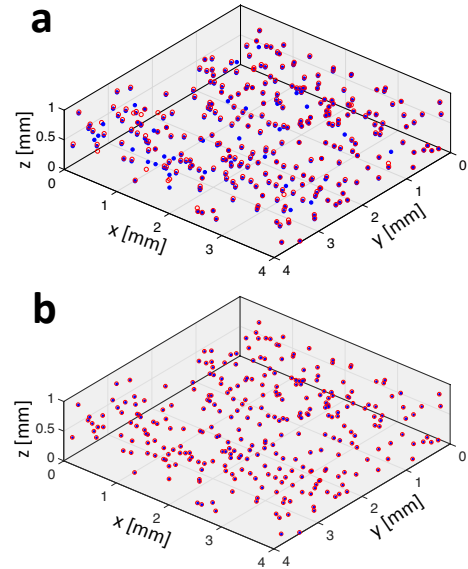

Figure 2. Numerical simulation results for the particle detection for different reconstruction resolution, (a) 0.1 mm and (b) 0.01 mm.

## 2. Shallow neural network model for mapping functions

In this study, we will consider a problem that has a liquid-vapor interface such as an evaporating droplet. In this case, an optical distortion occurs due to the refractive index jump. Furthermore, if a flow structure is complicated inside the liquid phase, typical mapping functions can be used as Equation (1) to reconstruct three-dimensional particle location for the two-camera system [21].

$$\begin{aligned}
x_1 &= a_{10} + a_{11}X + a_{12}Y + a_{13}Z + a_{14}XY + a_{15}YZ \\
&\quad + a_{16}XZ + a_{17}X^2 + a_{18}Y^2 + a_{19}X^2 \\
y_1 &= b_{10} + b_{11}X + b_{12}Y + b_{13}Z + b_{14}XY + b_{15}YZ \\
&\quad + b_{16}XZ + b_{17}X^2 + b_{18}Y^2 + b_{19}X^2 \\
x_2 &= a_{20} + a_{21}X + a_{22}Y + a_{23}Z + a_{24}XY + a_{25}YZ \\
&\quad + a_{26}XZ + a_{27}X^2 + a_{28}Y^2 + a_{29}X^2 \\
y_2 &= b_{20} + b_{21}X + b_{22}Y + b_{23}Z + b_{24}XY + b_{25}YZ \\
&\quad + b_{26}XZ + b_{27}X^2 + b_{28}Y^2 + b_{29}X^2
\end{aligned} \quad (1)$$

where $(x_i, y_i)$ is the location of a particle in $(X, Y, Z)$ cartesian coordinates and the index $i$ is for the number of the camera (here, $i$ is 1 and 2). $a_{ij}$ and $b_{ij}$ are coefficients of each mapping function where $j = 0, 1, 2, \ldots,$ and 9. The coefficients are obtained from the calibration and $(x_1, y_1)$ and $(x_2, y_2)$ are obtained from two cameras. The three-dimensional particle position is estimated from the best matched point among four curved planes generated from Equation (1). To do this, it requires multiple iterations to find the best position based on given criteria. Alternatively, a shooting method is used that arbitrary three-dimensional values $(X, Y, Z)$ keep inserting into the mapping functions until satisfying Equation (1). Likewise, this iterative conventional method increases the amount of calculation time as the mapping functions are complicated and the number of particles is increased.

Now, we propose a shallow neural network (SNN) model to reduce the computation time for the conventional mapping functions represented in Equation (1). First, the centroid of particles was detected to obtain the coordinate $(x_{I\_k}, y_{I\_k})$. We used the in-house function in MATLAB 'imfindcircles' function which finds all of the circles in a range of radius. The machine-learning network, the SNN model based on a built-in MATLAB function, consists of inputs, hidden layer, output layer, and outputs where we properly modified the MATLAB function for our study from Deep Learning Toolbox of MATLAB. The hidden layer is defined as $A = \sigma[W_h(x_1, y_1, x_2, y_2) + B_h]$ where $W_h$ is weighting coefficients and $B_h$ is bias. The matrix $A$ is an outcome of the hidden layer, which is transformed in terms of sigmoid functions [22]. The result from the hidden layers updated and inserted into the output layer, resulting in three-dimensional particle positions we obtained through the following equation, $AW_o + B_o$ where $W_o$ is a weighting coefficient and $B_o$ is a bias for the output layer. The trained results are examined by comparing with true values, i.e. the actual three-dimensional particle location. Here, we used Levenberg-Marquardt algorithm to train the results [23, 24] and initial coefficients and variables are set as default values of the built-in MATLAB function. In this procedure, 10 counts of hidden neurons and 3 counts of output neurons are contained in the converted mapping function. The training results iterate until perfectly matching between estimation and true value based on the regression analysis. Here, we set that 70 % of input data are trained as a training set, 15 % of the data are used as a validation set and the remained data are examined as a test set. From this training process, the weighting coefficients and bias variables are updated every single step.

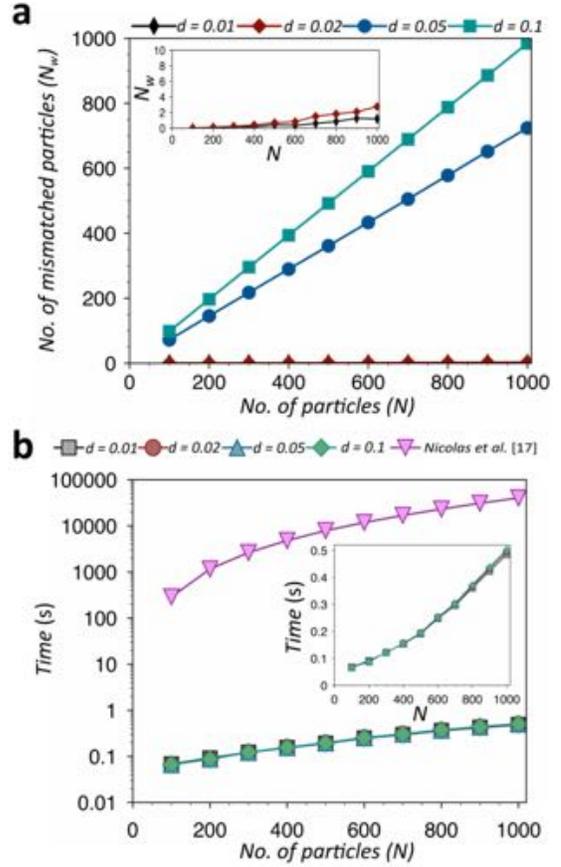

Figure 3. Effects of number of particles and reconstruction mesh resolutions on the particle reconstruction and the computation time. (a) Particle reconstruction results by varying the total number of particles. (b) Comparison of calculation time for different reconstruction mesh resolution $d$, i.e. $d = 0.1, 0.05, 0.02$ and $0.01$ mm.

After the machine-learning process, the three-dimensional particle location $(X, Y, Z)$ is calculated through the mapping function based on SNN model, $F_{nn}(x_1, y_1, x_2, y_2)$, which is described in Figure 1, where there are 10 hidden layers for the machine learning. To confirm the accuracy of the particle reconstruction, the error $(\varepsilon)$ of the trained results is evaluated using following equations,

$$\begin{aligned}
\varepsilon_{x_1} &= \left| \begin{array}{l} x_1 - (a_{10} + a_{11}X + a_{12}Y + a_{13}Z + a_{14}XY \\ +a_{15}YZ + a_{16}XZ + a_{17}X^2 + a_{18}Y^2 + a_{19}Z^2) \end{array} \right|, \\
\varepsilon_{y_1} &= \left| \begin{array}{l} y_1 - (b_{10} + b_{11}X + b_{12}Y + b_{13}Z + b_{14}XY \\ +b_{15}YZ + b_{16}XZ + b_{17}X^2 + b_{18}Y^2 + b_{19}Z^2) \end{array} \right|, \\
\varepsilon_{x_2} &= \left| \begin{array}{l} x_2 - (a_{20} + a_{21}X + a_{22}Y + a_{23}Z + a_{24}XY \\ +a_{25}YZ + a_{26}XZ + a_{27}X^2 + a_{28}Y^2 + a_{29}Z^2) \end{array} \right|, \\
\varepsilon_{y_2} &= \left| \begin{array}{l} y_2 - (b_{20} + b_{21}X + b_{22}Y + b_{23}Z + b_{24}XY \\ +b_{25}YZ + b_{26}XZ + b_{27}X^2 + b_{28}Y^2 + b_{29}Z^2) \end{array} \right|, \quad (2)
\end{aligned}$$

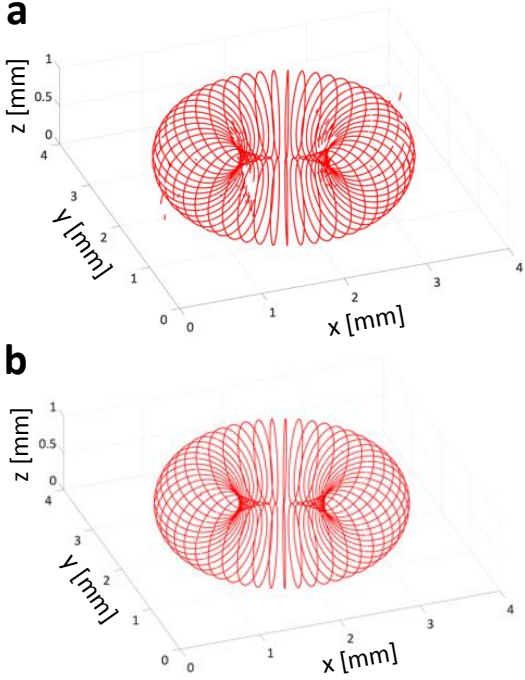

Figure 4. Numerical simulation results from Equation (4) for particle detection with different reconstruction resolutions, (a) 0.1 mm and (b) 0.01 mm.

where $\varepsilon_{x_1}, \varepsilon_{y_1}, \varepsilon_{x_2}$ and $\varepsilon_{y_2}$ represent the reliability of particle connection, relatively. Based on this, the total error ($\varepsilon_{tot}$) is defined as RMS (root-mean-square) deviation, such as $\varepsilon_{tot} = \sqrt{\varepsilon_{x_1}^2 + \varepsilon_{y_1}^2 + \varepsilon_{x_2}^2 + \varepsilon_{y_2}^2}$. Among the set of errors, the minimal one is compared with a relative error ($\varepsilon_m$), which is set as 5 pixels that is the approximated individual particle image diameter.

In order to evaluate the suggested algorithm, we performed a numerical simulation. To do this, artificial particles are randomly generated in a finite volume, i.e. $4 \times 4 \times 1$ mm$^3$. The number of test particles ranges from 100 to 1000 and we also vary the reconstruction mesh resolution, e.g. $d$ = 0.01, 0.02, 0.05, and 0.1 mm. Here, the mesh resolution indicates the minimal window size for the particle reconstruction in a volume. The reconstruction mesh resolution defines a training interval in simulation space when creating neural network fitting function $F_{nn}$. To test our method, we calculate the reconstruction error between trained results and true values based on the following equation,

$$\varepsilon_{dis} = \frac{\sqrt{(X_{true}-X')^2 + (Y_{true}-Y')^2 + (Z_{true}-Z')^2}}{h}, \quad (3)$$

where $h$ is the height of the target volume, $(X_{true}, Y_{true}, Z_{true})$ and $(X', Y', Z')$ are three-dimensional positions of pre-determined particles (i.e., true values) and calculated particles with trained results, respectively. Each numerical simulation was 100 times repeated in order to confirm reproducibility. In this work, we set the particle matching criterion, $\varepsilon_{dis} < 10^{-3}$.

Table 1. Numerical simulation results for particle detection for different reconstruction resolution. The results show computation time, mismatched particles, and velocity.

| Reconstruction resolution [mm] | 0.1 | 0.05 | 0.02 | 0.01 |
|---|---|---|---|---|
| Computation time [s] | 6.208 | 6.343 | 6.170 | 6.157 |
| Mismatched particles [count] | 5.668 | 1.094 | 1.046 | 1.064 |
| Velocity [μm/s] | 6.6 | 6.6 | 6.5 | 6.5 |

## 3. Numerical simulation

The numerical results of particle detection according to the mesh size (reconstruction resolution, $d$) are presented in Figure 2. In this case, the total number of reference particles is 300 counts that are randomly generated using MATLAB functions. Blue dots and red circles indicate the reference values and estimated particles, respectively. As shown in Figures 2 and 3, the mesh resolution increases with the particle detection rate. The coefficients of inverse mapping function from the $F_{nn}$ can be generated more accurately as the mesh resolution gets smaller due to more training sets. If the mesh size for the reconstruction is smaller than 1 % of the height, a mismatching rate is almost negligible, < 0.1 %, as shown in Figure 3(a). To compute the results, we used the computer, Intel® Core™ i5-7500 CPU @ 3.40GHz and 16.0GB RAM. Here, in the case of the smallest mesh size and 10$^3$ particles, it only takes 0.5 second. We observed that in the current method the total computation time does not depend on the mesh resolution much in given conditions if SNN is applied, as shown in Figure 3(b). On the other hand, for the conventional 3D-PTV algorithm based on the gradient descent method [18], it takes around 11 hours for 1000 particles, as shown in Figure 3(b).

The numerical simulation was performed to reconstruct not only randomly positioned particles but also the unique flow pattern. The chosen numerical model features the rotating particles from the origin making the ring trajectories with time. The total number of particles is set 50, making one full rotation with 500 frames. The model is created based on the following conditions,

$$x = a \sin \theta' \times \sin \theta + C(x) + 2,$$
$$y = a \sin \theta' \times \cos \theta + C(y) + 2,$$
$$z = b \cos \theta' + C(z),$$
$$C(x) = r \sin \theta,$$
$$C(y) = r \cos \theta,$$
$$C(z) = 0.5, \quad (4)$$

where $a$, $b$ and $r$ represent the major and minor axis of ellipse and radius of ring. $\theta$ represents the angle of the particle between 0 and $2\pi$ with uniform distribution. $\theta'$ represents the angle of the rotating trajectories of the ring.

We also varied the reconstruction mesh resolution, e.g. 0.01, 0.02, 0.05, and 0.1 mm. In Figure 4, we present that the reconstruction result of the model according to the different mesh size, such as 0.1 and 0.01 mm. The permissive error was defined as $\varepsilon_{per} = \sqrt[3]{\frac{P_x \times P_y \times P_z}{n}}$ where $P_x, P_y,$ and $P_z$ represent the resolution of each image and $n$ represents the number of particles. It can be interpreted as the average length of uniformly distributed particles in the reconstruction space and this error means the effectiveness of the simulation. The minimum standard was set 0.05 (5%), which can vary in situation. The results of $\varepsilon_{per}$ on our simulation are 0.032 for the most efficient condition.

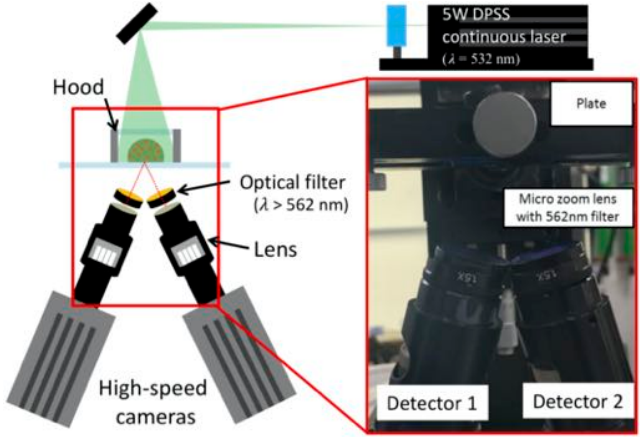

Figure 5. Schematic and experimental setting of stereoscopic three-dimensional particle tracking velocimetry (3D-PTV).

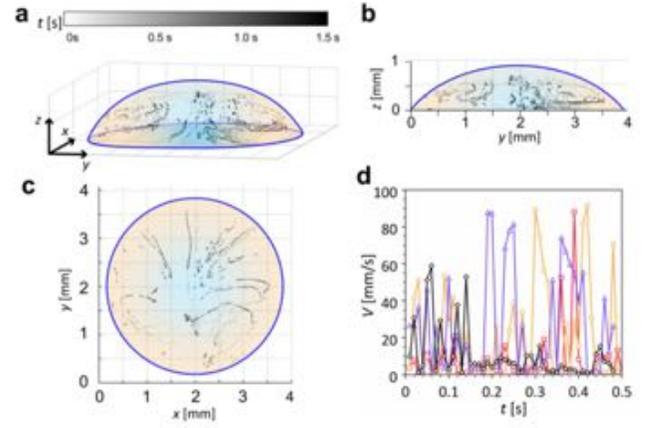

Figure 6. 3D-PTV results using SNN model for the complicated mixing flow pattern where the mixture droplet is 50 wt. % ethanol in distilled water. (a-c) Results of three-dimensional trajectories of particles inside the 5 μl binary mixture droplet during evaporation. (d) Time evolution of the velocity magnitude of multiple particles, where the different color indicates the different particle's velocity profile.

The result of simulation is summarized in Table 1. The computation time is extremely reduced compared with that of the conventional algorithm gradient descent methods [20], for 76,807 seconds, almost 21 hours. The mismatched particles on 500 frames are nearly 1 count except the resolution of 0.1 mm, over the 97.8 % of the particle pairing. The velocity calculated from the neural network fitting function was measured as perfectly matched with the reference velocity, 6.5 μm/s. From two numerical simulations, we believe that our SNN model can be used for the experiments with a high accuracy.

## 4. Experiments

Based on the SNN model, we performed the 3D-PTV experiments using the stereo-camera system, as shown in Figure 5. Using this setup, we investigated the solutal Marangoni flow inside an evaporating binary mixture drop. In the literature, it is reported that the binary mixture droplet initially has a complicated mixing flow and then it suddenly shows a radially outward flow during evaporation [25, 26]. However, in the meantime, most of results are two-dimensional flow field measurements. The randomly occurred three-dimensional solutal Marangoni flow has never been measured yet.

In this study, we tried to obtain the three-dimensional particles trajectories (also flow patterns) inside evaporating binary droplets using 3D-SNN-PTV. To perform 3D-PTV, two high-speed cameras having spatial resolution of 512 × 512 pixels with 100 frames per second (fps) are installed under the transparent glass plate (see Figure 5). A 5 μl volume drop of the binary mixture was placed on the glass plate. To capture the fluid motion, the 5 μm Rhodamine-B fluorescent particles (PS-FluoRed 5.0, Microparticles GmbH, Germany) are used, which is illuminated by 5W DPSS continuous laser (λ = 532 nm, Intech System) with an optical beam expander was used. We directly counted the seeding density of particles that is around 0.004 ppp (particles per pixel) in the current experiments. The particle image has approximately 5 pixels. The particle concentration is about 0.03 % in volume fraction. For this PTV, the Stokes number (St) is approximately $10^{-6}$, where $St = \tau U/R$; $\tau = (\Delta \rho d_p^2)/(18\mu)$ ; $U$ is the flow speed; $d_p$ is the particle diameter, $\Delta \rho$ is the density difference between fluid and particle, $R$ is the radius of the drop, and $\mu$ is the dynamic viscosity of the mixture. Thus, the particles can be assumed to follow the fluid motion almost perfectly [27]. Before the actual measurement, to obtain the mapping functions the calibration is done. Due to the liquid-vapor interface, the liquid was filled between the calibration target and the substrate and the target was dropped 0, 0.5, and 1 mm from a position of a glass plate. [28].

Using the current method, we measured full three-dimensional flow fields inside the binary mixture droplet during evaporation. At the initial stage, it shows continuous mixing flow patterns near the contact line, which are due to the surface tension difference by non-uniform solute concentration in the droplet due to the selective evaporation [25, 29]. The three-dimensional flow pattern is obtained using the suggested SNN method. As already mentioned in Figure 3(b), to reconstruct 1000 particles in a three-dimensional domain, the actual calculation of the current algorithm was done in about 0.5 s under the current computation power. Thus, to obtain one vector field in this experiment, one second is quite enough after execution, which means that we can use this algorithm for the real-time flow field measurement.

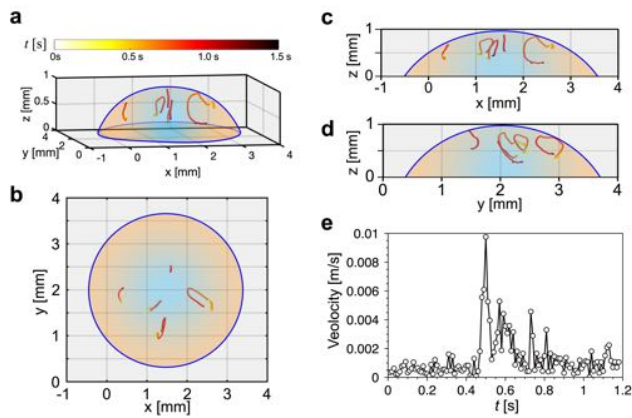

Figure 7. 3D-PTV results using SNN model for the bursting flow where the mixture droplet is 5 wt. % ethanol in distilled water. (a) 3D particles' trajectories inside a 5 μl ethanol- water droplet. (b), (c), and (d) are the top-view, side-view in the *x-z* plane, side-view in the *y-z* plane, respectively. (e) Time evolution of velocity of a single particle from PTV.

To examine the flow patterns during the evaporation of the binary mixture droplet, we focus on two parts; (1) complicated mixing flow regime and (2) random single bursting flow regime. First of all, we obtained the particles' trajectories inside the binary mixture droplet where the initial ethanol concentration in a water droplet is relatively high, e.g. 50 wt. % ethanol in distilled water, as shown in Figure 6. In the literatures, the typical flow speed was reported as $O(100$ μm/s). However, we obtain at the continuous mixing flow regime the typical flow speed ranges from $O(10$ mm/s) to $O(1$ mm/s) as shown in Figure 6(d), which is faster than the literature results. We believe that the literature results are the depth-averaged flow speed due to the conventional two-dimensional PIV, which is an underestimated result. From the results by the 3D SNN-PTV method, we monitor the speed of the individual particle, which is presented as a different color in Figure 6(d). We observed that the individual particles randomly accelerate and decelerate, which are due to the non-uniform surface tension distribution along the interface of the binary mixture droplet during evaporation.

If the concentration of ethanol in the binary mixture is relatively low, after the complicated flow pattern, the flow suddenly stops. After few seconds, we can observe the radially outward flow pattern that occurs nearly at the liquid-gas interface. For now, it is difficult to fully understand the reasons how the local Marangoni stress occurs in such a way and how it induces a bursting flow because it is almost impossible to visualize the ethanol molecule in the droplet and near the liquid-gas interface. Nevertheless, we successfully reconstruct the three-dimensional particle positions and measure the trajectories using 3D SNN-PTV method. Based on this, we could estimate the bursting flow speed in time, which occurs temporarily and randomly occurred. Figure 7 presents that the 3D particles' trajectories and the velocity profile of one of the particles in the droplet. To avoid particles' interactions in the droplet, we only seeded a few particles. Furthermore, the bursting flow occurs after the complicated mixing flow, so that the seeded particles are sedimented on the substrate. Thus, the detected number of particles are few. In Figure 7(e), we observed that the particle speed reaches to almost 0.01 m/s, which is also extremely fast compared to the previously measured data in [23]. In the literature, they performed conventional two-dimensional PIV and reported that the maximum speed of the bursting flow is $10^{-4}$ m/s, which is truly underestimated results.

## 5. Conclusions

In this paper, for three-dimensional particle position reconstruction for stereoscopic 3D-PTV, to reconstruct three-dimensional particle positions, we propose a machine-learning approach based on the SNN method, which is extremely fast compared to the conventional method. We show that this method can enhance computation efficiency with a high accuracy. We believe that the suggested algorithm can be applied for the real-time flow field measurement. Even further, if we could apply deep-learning algorithm convolutional neural network (CNN) in the near future, the efficiency of computation power and precision would be maximized.

**Acknowledgement**
National Research Foundation of Korea (NRF) (NRF-2019R1A2C2003176 and NRF-2018R1C1B6004190)